\documentclass[twocolumn]{article}
\usepackage[margin=.6in]{geometry}
\usepackage{graphicx}
\usepackage{amsmath}
\usepackage{amssymb}
\usepackage{booktabs} 
\usepackage{multirow} 
\usepackage{hyperref} 
\usepackage{algorithm} 
\usepackage{amsfonts} 
\usepackage{fontspec} 
\usepackage{threeparttable}
\usepackage[inline]{enumitem}
\usepackage{authblk}
\usepackage{natbib}

\DeclareMathOperator*{\softmax}{softmax}

\title{SPECTRE: Spectral Pre-training Embeddings with Cylindrical Temporal Rotary Position Encoding for Fine-Grained sEMG-Based Movement Decoding}

\author[1,2]{Zihan Weng\thanks{Equal contributions}}
\author[1,2]{Chanlin Yi\textsuperscript{*}}
\author[3]{Pouya Bashivan}
\author[1,2]{Jing Lu}
\author[1,2,4,5]{Fali Li}
\author[1,2,4,6]{Dezhong Yao}
\author[7]{Jingming Hou\thanks{\noindent Corresponding author: jingminghou@hotmail.com, zhangysacademy@gmail.com, xupeng@uestc.edu.cn}}
\author[8,9]{Yangsong Zhang\textsuperscript{\textdagger}}
\author[1,2,4,10,11]{Peng Xu\textsuperscript{\textdagger}}

\affil[1]{Clinical Hospital of Chengdu Brain Science Institute, MOE Key Lab for Neuroinformation, University of Electronic Science and Technology of China, Chengdu, China}
\affil[2]{School of Life Science and Technology, Center for Information in Medicine, University of Electronic Science and Technology of China, Chengdu, China}
\affil[3]{Department of Physiology, McGill University, Montreal, Canada}
\affil[4]{Research Unit of NeuroInformation, Chinese Academy of Medical Sciences, Chengdu, China}
\affil[5]{Department of Electrical and Computer Engineering, Faculty of Science and Technology, University of Macau, Macau, China}
\affil[6]{School of Electrical Engineering, Zhengzhou University, Zhengzhou, China}
\affil[7]{Department of Rehabilitation, Southwest Hospital, Army Medical University, Chongqing, China}
\affil[8]{Laboratory for Brain Science and Artificial Intelligence, School of Medicine, Southwest University of Science and Technology, Mianyang, China}
\affil[9]{School of Computer Science and Technology, Southwest University of Science and Technology, Mianyang, China}
\affil[10]{Radiation Oncology Key Laboratory of Sichuan Province, Chengdu, China}
\affil[11]{Rehabilitation Center, Qilu Hospital of Shandong University, Jinan, China}

\date{}

\begin{document}

\maketitle

\begin{abstract}
    Decoding fine-grained movement from non-invasive surface Electromyography (sEMG) is a challenge for prosthetic control due to signal non-stationarity and low signal-to-noise ratios. Generic self-supervised learning (SSL) frameworks often yield suboptimal results on sEMG as they attempt to reconstruct noisy raw signals and lack the inductive bias to model the cylindrical topology of electrode arrays. To overcome these limitations, we introduce SPECTRE, a domain-specific SSL framework. SPECTRE features two primary contributions: a physiologically-grounded pre-training task and a novel positional encoding. The pre-training involves masked prediction of discrete pseudo-labels from clustered Short-Time Fourier Transform (STFT) representations, compelling the model to learn robust, physiologically relevant frequency patterns. Additionally, our Cylindrical Rotary Position Embedding (CyRoPE) factorizes embeddings along linear temporal and annular spatial dimensions, explicitly modeling the forearm sensor topology to capture muscle synergies. Evaluations on multiple datasets, including challenging data from individuals with amputation, demonstrate that SPECTRE establishes a new state-of-the-art for movement decoding, significantly outperforming both supervised baselines and generic SSL approaches. Ablation studies validate the critical roles of both spectral pre-training and CyRoPE. SPECTRE provides a robust foundation for practical myoelectric interfaces capable of handling real-world sEMG complexities.
    \end{abstract}


\section{Introduction}\label{sec:introduction}
The human hand, with its remarkable dexterity, enables intricate interactions with the physical world, forming the cornerstone of countless daily activities and complex skills \citep{roda-salesElectromyographyKinematicsData2023}. Capturing the neural intent behind these fine-grained finger movements through non-invasive means is a critical pursuit in fields like advanced prosthetics \citep{jiangBioroboticsResearchNoninvasive2023}, neurorehabilitation \citep{zhangWearableMasterSlave2023}, and intuitive human-computer interfaces (HCIs) \citep{liuPracticalSystem3D2023}. Among various biosignals, surface Electromyography (sEMG), which measures the electrical activity produced by skeletal muscles, offers a promising window into motor intent, originating from neural commands and reflecting neuromuscular activation patterns \citep{farinaExtractionNeuralInformation2014}. Its non-invasive nature and potential for real-time decoding make it particularly attractive.

Despite its potential, decoding continuous, fine finger movements from sEMG presents substantial challenges. sEMG signals are notoriously non-stationary and susceptible to various noise sources, including muscle fatigue, electrode shift, skin impedance variations, motion artifacts, and environmental interference \citep{farinaInfluenceAnatomicalPhysical2002}. Furthermore, significant inter-subject and even intra-subject variability exists due to differences in anatomy, muscle activation strategies, and recording conditions. Traditional machine learning approaches \citep{daiFingerJointAngle2020} often struggle with the complexity and variability of sEMG for continuous, multi-degree-of-freedom (DoF) decoding.

Deep learning, particularly Convolutional Neural Networks (CNNs) and Recurrent Neural Networks (RNNs), has shown promise in improving sEMG-based gesture recognition and movement decoding \citep{guoMultiAttentionFeatureFusion2023}. However, these supervised methods typically require large amounts of accurately labeled data, where continuous finger kinematics are precisely synchronized with multi-channel sEMG recordings. Acquiring such datasets is laborious, expensive, and particularly difficult for individuals with limb differences \citep{ctrl-labsatrealitylabsGenericNoninvasiveNeuromotor2024}.

Self-supervised learning (SSL) has emerged as a powerful paradigm to learn robust representations from large unlabeled datasets, mitigating the reliance on manual annotation \citep{huang2021self}. Landmark models in natural language processing (e.g., BERT \citep{devlinBERTPretrainingDeep2019}) and computer vision (e.g., MAE \citep{heMaskedAutoencodersAre2021}) have demonstrated the power of masked signal modeling. However, the unique characteristics of sEMG present significant hurdles for the direct adaptation of these successful frameworks. Critically, sEMG signals are fundamentally different from language or images; they are highly stochastic, non-stationary, and their information content is encoded not just in temporal sequences but also in frequency-domain patterns and complex spatial relationships across muscles.

Applying generic SSL methods to sEMG often leads to suboptimal outcomes for two primary reasons. First, methods like Masked Autoencoders (MAE) that aim to reconstruct raw signal force the model to learn the structure of noise and artifacts, rather than the underlying, more stable physiological signals. The physiologically relevant information in sEMG, related to motor unit action potential firing rates, is often better captured in the spectral domain \citep{farinaExtractionNeuralInformation2014}. Second, existing models typically treat multi-channel sEMG as a simple set of independent time series or a 2D image, ignoring the crucial topological information of the sensor layout---for instance, the cylindrical arrangement of electrodes around a limb, which is key to understanding agonist-antagonist muscle synergies. A model lacking this geometric inductive bias will struggle to efficiently learn these vital spatial dependencies.

To address these specific challenges, we propose a novel SSL framework specifically designed for sEMG representation learning, i.e., Spectral Pre-training Embeddings with Cylindrical Temporal Rotary Position Encoding, termed SPECTRE. Our framework is built on the hypothesis that an effective sEMG representation must be learned by focusing on physiologically relevant spectral features while explicitly modeling the physical topology of the sensors. The SPECTRE introduces innovations to achieve this:

\textbf{Domain-Specific Spectral Pre-training Task:} We propose a novel SSL pretext task based on masked prediction of discrete pseudo-labels derived from K-means clustering of Short-Time Fourier Transform (STFT) representations. This approach deliberately abstracts away from the noisy raw signal, compelling the model to learn robust representations of motion-relevant spectral patterns.

\textbf{Topology-Aware Positional Encoding:} We introduce Cylindrical Rotary Position Embedding (CyRoPE), a novel extension of Rotary Position Embedding (RoPE) \citep{suRoFormerEnhancedTransformer2023} for multi-channel time-series data. CyRoPE factorizes positional information into a temporal (linear) component and a spatial/channel (annular) component, explicitly encoding the typical cylindrical geometry of forearm sEMG electrode arrays. This allows the Transformer model to more effectively capture spatio-temporal dependencies and muscle synergies.

\textbf{Optimized sEMG Transformer Architecture:} Our proposed SPECTRE employs a hybrid CNN-Transformer backbone, using CNNs for robust local feature extraction and noise resilience, coupled with architectural enhancements like SwiGLU activation function \citep{shazeerGLUVariantsImprove2020} and RMS Normalization \citep{zhangRootMeanSquare2019} known to improve training stability and performance in large transformer models.

\textbf{Comprehensive Empirical Validation:} We conduct extensive experiments on multiple sEMG datasets, including a large-scale dataset and an extra dataset from individuals with transradial amputation. Results demonstrate that SPECTRE significantly outperforms several state-of-the-art SSL baselines, i.e., MAE, BIOT, VQ-MTM, and supervised models in continuous fine finger movement decoding. Rigorous ablation studies further validate the crucial contributions of both the spectral pre-training task and CyRoPE.

Our work suggests that designing SSL strategies that explicitly incorporate domain knowledge—such as the relevance of spectral features and sensor topology for sEMG—is critical for unlocking the full potential of deep learning in complex biosignal analysis. SPECTRE sets a new benchmark for SSL in sEMG and offers a promising foundation for building more practical and effective myoelectric interfaces.

The remainder of this paper is structured as follows: Section \ref{sec:related_work} reviews related work in sEMG decoding, SSL, and positional encoding. Section \ref{sec:methodology} details the SPECTRE framework, including the architecture, CyRoPE, and the spectral pre-training task. Section \ref{sec:experiments} describes the experimental setup, datasets, baselines, and presents the results. Section \ref{sec:discussion} analyzes the findings, discusses implications and limitations. Section \ref{sec:conclusion} concludes the paper.

\section{Related Work} \label{sec:related_work}

\subsection{sEMG-based Movement Decoding}
The exploration of decoding movement intents from sEMG signals has a rich history. Early approaches focused on extracting hand-crafted features (e.g., time-domain statistics, frequency-domain power \citep{Phinyomark2012FeatureEA}) and feeding them into machine learning models like Linear Discriminant Analysis (LDA) \citep{daiFingerJointAngle2020}, Support Vector Machines (SVMs) \citep{tavakoliRobustHandGesture2018}, and Hidden Markov Models (HMMs) \citep{wenHumanHandMovement2021}. While successful for classifying discrete gestures, these methods often lack the capacity to accurately regress the continuous, multi-DoF kinematics necessary for fine motor control \citep{fangAttributeDrivenGranularModel2021}.

Deep learning has significantly advanced the field. CNNs excel at extracting spatial and short-term temporal patterns from multi-channel sEMG \citep{fajardoEMGHandGesture2021}, while RNNs (like LSTMs) handle longer temporal dependencies \citep{karnamEMGHandNetHybridCNN2022}. Hybrid CNN-RNN/LSTM models \citep{wangHandGestureRecognition2023} have shown strong performance. More recently, Transformer architectures, leveraging self-attention, have been explored for their ability to capture global context and long-range dependencies in sEMG \citep{linBERTBasedMethod2023, Weng2025Realtime}. \cite{Weng2025Realtime} demonstrated the effectiveness of a CNN-Transformer hybrid for fine finger decoding, forming a basis for the SPECTRE architecture. However, all these supervised approaches heavily depend on large labeled datasets.

\subsection{Self-Supervised Learning (SSL)}
SSL aims to learn useful representations from unlabeled data by defining pretext tasks that leverage the inherent structure in the data. Major paradigms include:

\textbf{Contrastive Learning:} It learns representations by pulling augmented views of the same sample closer while pushing views from different samples apart (e.g., SimCLR \citep{chen2020simple}, MoCo \citep{he2020momentum}).

\textbf{Masked Signal Modeling:} It predicts masked or corrupted portions of the input signal (e.g., BERT \citep{devlinBERTPretrainingDeep2019} for text, MAE \citep{heMaskedAutoencodersAre2021} for images, HuBERT \citep{hsuHuBERTSelfSupervisedSpeech2021} for speech).

\textbf{Clustering-based Methods:} It groups similar instances together and uses cluster assignments as pseudo-labels (e.g., DeepCluster \citep{caron2018deep}, SwAV \citep{caron2020unsupervised}).

\textbf{Redundancy Reduction:} It minimizes redundancy between different views of the same sample (e.g., Barlow Twins \citep{zbontar2021barlow}).

SPECTRE primarily builds upon the masked signal modeling paradigm but introduces a novel target representation based on clustered spectral features, specifically designed for sEMG signals.

\subsection{SSL for Biosignals}
SSL is gaining traction in biosignal processing, notably for EEG and audio/speech signals. The contrastive learning \citep{Yi2023LearningTE} and masked prediction have been explored in EEG researches \citep{chienMAEEGMaskedAutoencoder2022, kostas2021bendr}. \cite{yang2023biot} proposed a general Transformer for various biosignals using masked prediction and channel dropout. \cite{gui2024vector} applied vector quantization to masked time-series modeling for EEG, drawing parallels with speech models like HuBERT \citep{hsuHuBERTSelfSupervisedSpeech2021} that predict discrete acoustic units. These methods often focus on raw signal reconstruction or generic token prediction. By contrast, SPECTRE specifically focuses on sEMG challenges and utilizes STFT-derived spectral features, grounded in EMG physiology, for generating pseudo-labels, rather than relying on raw signals. Besides, we introduce CyRoPE for explicit spatio-temporal modeling tailored to electrode array topology. Previous SSL applications targeting continuous sEMG regression are limited, and often apply generic frameworks such as Masked Autoencoder (MAE) \citep{heMaskedAutoencodersAre2021} or basic contrastive learning without the domain-specific adaptations proposed in SPECTRE.

\subsection{Positional Encoding in Transformers}
Standard Transformers are permutation-equivariant and require positional encoding (PE) to incorporate sequence order. Early approaches included sinusoidal or learned absolute PEs \citep{vaswaniAttentionAllYou2017}. Relative PEs encode pairwise positional relationships \citep{shaw2018selfattention, raffel2020exploring}. The RoPE \citep{suRoFormerEnhancedTransformer2023} has emerged as a highly effective technique, particularly in large language models \citep{grattafiori2024llama3herdmodels, qwen2025qwen25technicalreport}. RoPE applies rotations to query and key vectors based on their absolute positions, such that their dot product in the attention mechanism inherently depends only on their relative positions. While powerful, standard RoPE is one-dimensional. For multi-channel sEMG, where signals have both temporal and spatial (channel) dimensions, 1D PE is insufficient. CyRoPE addresses this limitation by extending RoPE to factorize and encode both the linear temporal progression and the annular spatial arrangement characteristic of sEMG sensor placements. This explicit encoding of the sensor topology provides a strong inductive bias for learning spatio-temporal patterns.

\section{Methodology} \label{sec:methodology}

\begin{figure*}[!htbp]
    \centering
    \includegraphics[width=\textwidth]{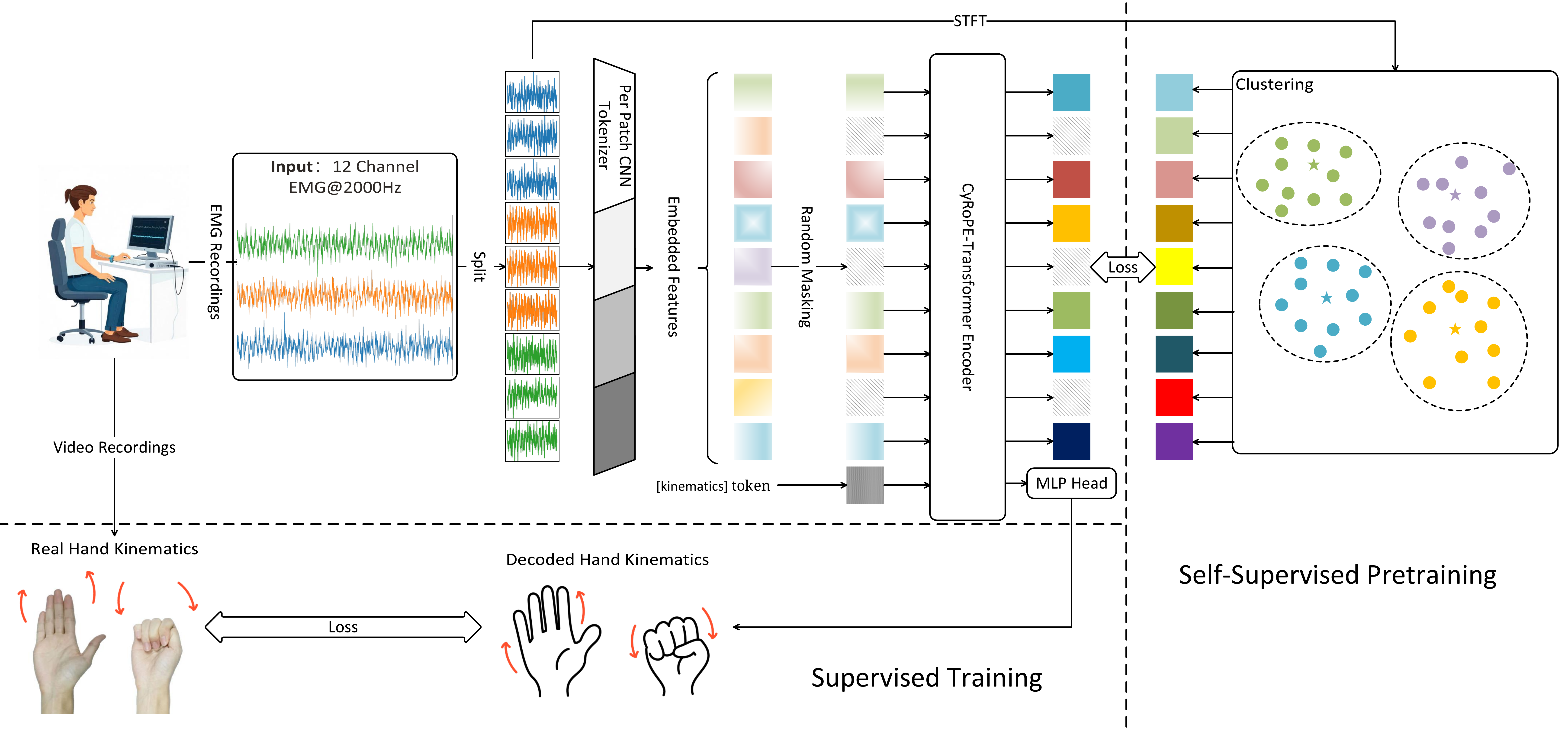} 
    \caption{Schematic overview of the SPECTRE framework. Raw multi-channel sEMG  is first processed by a channel-independent CNN embedding layer to produce spatio-temporal patches . These patches are fed into a Transformer Encoder incorporating CyRoPE. During pre-training (Top Right), a random subset of patches is masked. The encoder processes unmasked patches. The pre-training objective is to predict clustered STFT pseudo-labels  corresponding to the masked patches using a prediction head. During fine-tuning, a [kinematics] or kinematic token is often used, and its output representation from the Transformer is passed to an MLP head to predict the downstream task target, e.g., finger joint angles.}
    \label{fig:model-full-schematic}
\end{figure*}

In this section, we introduce SPECTRE, a self-supervised learning framework tailored for extracting robust representations from multi-channel sEMG signals. Our design philosophy is guided by the need to address the dual challenges of local temporal feature extraction under high noise and modeling global spatio-temporal muscle synergies. To this end, SPECTRE's architecture (Fig. \ref{fig:model-full-schematic}) is a synergistic hybrid of a Convolutional Neural Network (CNN) front-end and a Transformer back-end.

Given a raw multi-channel sEMG input $X \in \mathbb{R}^{C \times L}$, where $C$ is the number of channels and $L$ is the sequence length, the processing pipeline is as follows:

\textbf{Channel-Independent CNN Embedding:} The input $X$ is first processed by a CNN module applied independently to each channel. This extracts local temporal features, performs initial dimensionality reduction, and segments the signal into patches. This yields a sequence of patch embeddings $Z \in \mathbb{R}^{N \times d}$, where $N = C \times (L/P)$ is the total number of patches (assuming patch length $P$ and stride $P$), and $d$ is the embedding dimension.

\textbf{Positional Encoding (CyRoPE):} Our novel CyRoPE mechanism is applied to the patch embeddings $Z$ to inject spatio-temporal position information reflecting both time progression and channel arrangement.

\textbf{Transformer Encoder:} The positionally-encoded embeddings are processed by a stack of Transformer encoder layers. Each layer comprises multi-head self-attention (MSA) and a feed-forward network (FFN). We utilize RMSNorm \citep{zhangRootMeanSquare2019} for layer normalization and SwiGLU activation function \citep{shazeerGLUVariantsImprove2020} within the FFN for improved performance and stability.

\textbf{Prediction Head:} Depending on the operation phase (pre-training or fine-tuning), a specific prediction head operates on the Transformer output representations.

\subsection{Channel-Independent CNN Embedding}
To effectively process multi-channel sEMG signals while preserving local temporal information and managing computational complexity for the Transformer, we utilize a channel-independent CNN embedding strategy.
The input $X \in \mathbb{R}^{C \times L}$ is first divided into non-overlapping patches along the time dimension for each channel independently. Let the patch length be $P$. This results in $L/P$ patches per channel, totaling $N = C \times (L/P)$ patches. Each patch $p_{c,t} \in \mathbb{R}^{P}$ (where $c$ is the channel index, $1 \le c \le C$, and $t$ is the temporal patch index, $1 \le t \le L/P$) retains its original channel and temporal location information.
Each patch $p_{c,t}$ is then processed by a shared CNN encoder $f_{cnn}: \mathbb{R}^{P} \to \mathbb{R}^{d}$ to produce the embedding $z_{c,t} = f_{cnn}(p_{c,t})$.
The CNN encoder $f_{cnn}$ consists of multiple 1D convolutional layers with  GELU activation function and pooling operation to reduce temporal resolution within the patch while extracting features.
This channel-independent processing allows the CNN to learn temporal patterns specific to muscle activity captured by individual channels before the Transformer models cross-channel interactions. The resulting sequence $Z = \{z_{c,t}\}$ forms the input tokens for the Transformer.

\subsection{Cylindrical Rotary Position Embedding (CyRoPE)} \label{subsec:cyrope}
Standard Transformer architectures are permutation-equivariant and require an external mechanism to understand sequence order. While various positional encoding (PE) schemes exist, they are ill-suited for multi-channel sEMG data from wearable devices. Absolute PEs struggle to generalize, while standard relative PEs, including the one-dimensional Rotary Position Embedding (RoPE) \citep{suRoFormerEnhancedTransformer2023}, treat the input as a simple linear sequence. This fails to capture the inherent spatio-temporal structure of sEMG, where signals are recorded simultaneously from electrodes arranged with a specific physical topology---often a cylindrical (annular) configuration around a limb. This spatial arrangement is not arbitrary; it reflects the underlying musculoskeletal anatomy, and encoding it is critical for modeling muscle synergies.

\begin{figure}[!htbp]
    \centering
    \includegraphics[width=\linewidth]{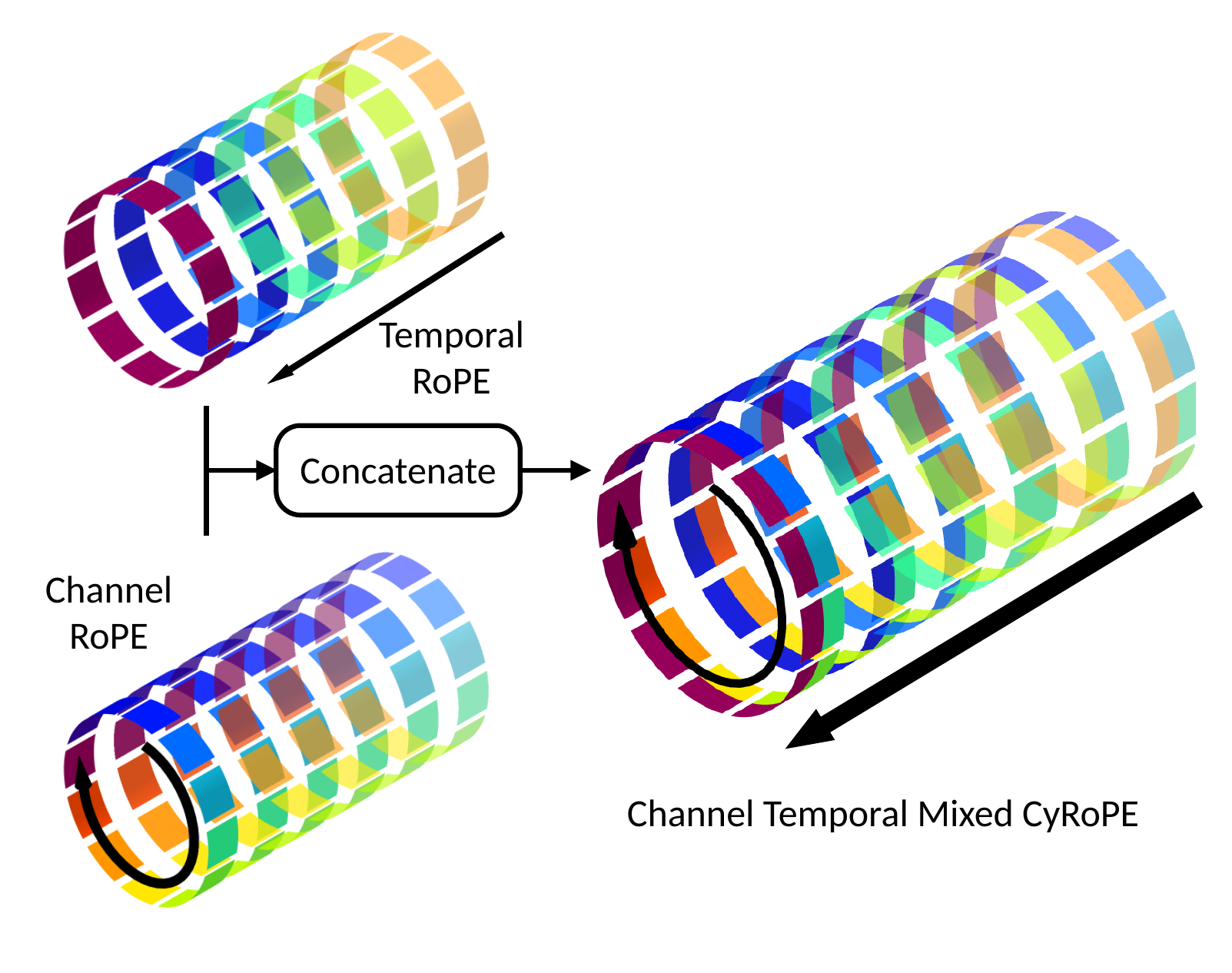}
    \caption{Schematic diagram of CyRoPE. In the two left figures, different colors represent different indices in the time and spatial dimensions. The time dimension and spatial dimension (EMG channel index) are first embedded separately using rotary position embedding, then merged into a single CyRoPE that simultaneously contains both temporal and spatial position information.}
    \label{fig:2drope}
\end{figure}

To address this limitation, we propose Cylindrical Rotary Position Embedding (CyRoPE), a principled extension of RoPE designed to inject geometric inductive bias into the self-attention mechanism. CyRoPE's core idea is to factorize the spatio-temporal positional information into two orthogonal components: a \textbf{linear temporal progression} and a \textbf{cylindrical spatial arrangement}, as illustrated in Fig. \ref{fig:2drope}. This factorization allows the model to learn relative positions in time and space simultaneously and distinctly.

Let $z \in \mathbb{R}^d$ be an embedding corresponding to temporal patch index $t$ and channel index $c$. We partition the embedding dimension $d$ into two equal halves: $d/2$ for temporal encoding and $d/2$ for spatial encoding. Let $z = [z_t \| z_c]$, where $z_t, z_c \in \mathbb{R}^{d/2}$. We apply rotary embeddings independently to $z_t$ and $z_c$.

Following RoPE, we view $\mathbb{R}^{d/2}$ as $\mathbb{R}^{d/4 \times 2}$. For the temporal part $z_t$, the rotation for position $t$ is defined by a block diagonal matrix $R_{\Theta_t}$:
\begin{equation}
    R_{\Theta_t} = \text{diag}(R_{\theta_{t,1}}, R_{\theta_{t,2}}, \dots, R_{\theta_{t,d/4}})
\end{equation}
where each $R_{\theta_{t,i}}$ is a 2D rotation matrix:
\begin{equation}
    R_{\theta_{t,i}} = \begin{pmatrix} \cos(t \theta_{t}^{(i)}) & -\sin(t \theta_{t}^{(i)}) \\ \sin(t \theta_{t}^{(i)}) & \cos(t \theta_{t}^{(i)}) \end{pmatrix}
\end{equation}
The frequencies are $\theta_t^{(i)} = 1 / \beta_t^{2i / (d/2)}$ for $i=1, \dots, d/4$, with a base $\beta_t$.
Here, we uses \(\beta_t = 10^4\), similar to traditional RoPE settings, while also slightly lowered to match the shorter actual length of myoelectric sequences\citep{kexuefm-10122}.
The rotated temporal embedding is $z'_t = R_{\theta_t} z_t$.

For the spatial/channel part $z_c$, we apply a similar rotation $R_{\Theta}$ based on the channel index $c$. The key innovation lies in defining the spatial rotation frequencies $\theta_{c}^{(i)}$ to reflect the cylindrical geometry. Assuming $C$ channels are arranged approximately equidistantly around a $2\pi$ circumference, the fundamental angular frequency corresponding to one full cycle across the channels is $\omega_0 = 2\pi / C$ radians per channel index step.

We anchor the RoPE frequency spectrum to this physical arrangement by setting the frequency for the lowest-frequency pair (conventionally $i=d/4$) to this fundamental physical frequency: $\theta_c^{(d/4)} = \omega_0$. In the standard RoPE formulation, $\theta_c^{(d/4)} = 1 / \beta_c^{2(d/4) / (d/2)} = 1/\beta_c$. Therefore, we set the spatial base $\beta_c$ as:
\begin{equation} \label{eq:beta_c_definition}
    \beta_c = \frac{1}{\omega_0} = \frac{C}{2\pi}
\end{equation}
The full set of spatial frequencies is derived using this base:
\begin{equation} \label{eq:phi_i_definition}
    \theta_c^{(i)} = \frac{1}{\beta_c^{2i / (d/2)}} = \omega_0^{2i / (d/2)} = \left( \frac{2\pi}{C} \right)^{2i / (d/2)}
\end{equation}
for $i=1, \dots, d/4$.

This choice ensures that the frequency for the pair $i=d/4$ is $\phi_{d/4} = \omega_0 = 2\pi/C$. Consequently, the rotation component $R_{\phi_{c,d/4}}$ encodes the absolute angular position $c \cdot \phi_{d/4} = c \cdot (2\pi/C)$ of the channel on the cylinder (assuming channel 0 corresponds to angle 0). The relative angles encoded by RoPE for this component, $(c_1 - c_2) \phi_{d/4} = (c_1 - c_2) (2\pi/C)$, directly reflect the physical angular separation between channels $c_1$ and $c_2$. The other frequencies $\phi_i$ capture relative spatial relationships at different, geometrically derived scales. This formulation makes the spatial encoding more explainable by directly linking it to the physical arrangement of the channels and enable interpolation to different channel configurations \citep{kexuefm-9675}.

The final CyRoPE-encoded embedding $z' = [z'_t \| z'_c]$ thus equips the Transformer with a native understanding of both "when" and "where" an sEMG signal occurs, respecting the linear nature of time and the cylindrical nature of the spatial layout.

The final CyRoPE-encoded embedding is $z' = [z'_t \| z'_c]$. When computing self-attention between a query $q$ at $(t_1, c_1)$ and a key $k$ at $(t_2, c_2)$, the application of these rotations ensures that the attention score $q^T k$ implicitly depends on the relative temporal distance $(t_1 - t_2)$ and the relative angular spatial separation encoded by $(c_1 - c_2)$ via the spatial frequencies $\phi^{(i)}$. This geometric inductive bias helps the model learn spatio-temporal patterns and muscle synergies that respect the physical layout of the sensors.

\subsection{Self-Supervised Pre-training: Masked Spectral Embedding Prediction} \label{subsec:ssl_task}
The core of SPECTRE's self-supervised learning approach is a pretext task designed to leverage the informative frequency-domain characteristics of sEMG, which are linked to muscle physiology (fiber type, contraction level, fatigue) \citep{4121265} and are often more robust to certain types of noise than raw time-domain signals.

The process involves the following steps:
\subsubsection{Masking} Following the MAE paradigm \citep{heMaskedAutoencodersAre2021}, we randomly mask a significant fraction (e.g., 30-50\%) of the input patch embeddings $Z$. Let $Z_{vis}$ denote the set of visible (unmasked) patches and $Z_{mask}$ the set of masked patches. Only $Z_{vis}$ (with CyRoPE applied) are passed as input to the Transformer encoder.

\subsubsection{STFT Pseudo-Label Generation} This step generates the target labels for the masked patches:
\begin{enumerate*}
    \item For each patch $p_{c,t}$ corresponding to an embedding $z_{c,t}$, extract the corresponding STFT representation across the patch duration. Let this be $S_{p_{c,t}} \in \mathbb{R}^{F \times T_{patch}}$, where $F$ is the number of frequency bins and $T_{patch}$ is the number of STFT frames within the patch.

    \item Flatten $S_{p_{c,t}}$ into a vector $s_{p_{c,t}} \in \mathbb{R}^{F \times T_{patch}}$.
    \item Offline collect these flattened spectral vectors $s_{p}$ from a large corpus of unlabeled sEMG data and apply K-means clustering to obtain $K$ cluster centroids $\{\mu_1, ..., \mu_K\}$.
    \item For each patch $p_{c,t}$ in the current pre-training batch, find its nearest cluster centroid $\mu_k$. This centroid $\mu_k$ serves as the pseudo-label target $\hat{y}_{SSL}^{(c,t)}$ for the patch $(c,t)$.
\end{enumerate*}

\subsubsection{Prediction Head} A lightweight linear MLP prediction head $f_{pred}: \mathbb{R}^d \to \mathbb{R}^{Dim(\mu_k)}$ is applied to the Transformer's output representations corresponding only to the masked positions. Let $h_{mask}$ be the output hidden states for masked patches.
\subsubsection{Pre-training Objective} The model is trained to minimize the reconstruction loss between the predicted spectral representations and the target cluster centroids for the masked patches:
\begin{equation}
    \mathcal{L}_{SSL} = - \frac{1}{|Z_{mask}|} \sum_{(c,t) \in \text{masked}} \log P( \hat{y}_{SSL}^{(c,t)} | h_{mask}^{(c,t)} )
    \label{eq:ssl_loss}
\end{equation}
where $P(k | h) = (\softmax(f_{pred}(h)))_k$ is the predicted probability of cluster $k$ given the hidden state $h$.

This spectral prediction task encourages the model to learn representations that capture discriminative frequency patterns related to muscle activity. The clustering step discretizes the spectral space, providing stable, denoised targets and potentially simplifying the learning task compared to regressing continuous spectral values or noisy raw signals.

\subsection{Fine-tuning for Movement Decoding}
After self-supervised pre-training, the learned weights of the SPECTRE encoder serve as a strong initialization for downstream tasks. For fine-grained movement decoding:
\begin{itemize}
    \item The input to $f_{ft}$ can be the representation of a special [kinematics] token prepended to the input sequence.
    \item A task-specific MLP head $f_{ft}: \mathbb{R}^d \to \mathbb{R}^{DoF}$ is added on top of the Transformer encoder to decode the [kinematics] token. $DoF$ is the number of finger movements/angles to predict, specifically, we set it to be 5 for five fingers' flexion value, respectively.
    \item The entire model is supervised fine-tuned on a labeled dataset $(X_j, Y_j)$, where $Y_j \in \mathbb{R}^{DoF}$ represents the ground truth kinematics, corresponding to the sEMG input $X_j$.
    \item The fine-tuning objective is Mean Squared Error (MSE): $\mathcal{L}_{FT} = \frac{1}{M} \sum_{j=1}^{M} || \hat{Y}_j - Y_j ||^2_2$ where $M$ is the number of labeled samples, and $h_j$ is the input representation to the fine-tuning head for sample $j$.
\end{itemize}

\section{Experiments} \label{sec:experiments}

We conducted extensive experiments to evaluate SPECTRE's effectiveness for continuous fine finger movement decoding. We compared its performance against relevant baselines, assessed the impact of pre-training data scale and domain, and further evaluated its applicability on amputation data, and performed ablation studies to isolate the contributions of its core components.

\subsection{Datasets}

We utilized several sEMG datasets representing different recording technologies, subject populations, and data volumes to ensure a comprehensive evaluation. All sEMG data was sampled at 2000 Hz.

\subsubsection{Basic flexible electrod dataset (Flex-Basic dataset)}
This dataset, described in \citep{Weng2025Realtime}, contains synchronized 12-channel sEMG signals recorded using custom flexible electrode sleeves and 5-DoF finger flexion kinematics tracked via video. Data was collected from 36 individuals without amputation subjects performing prompted finger flexion/extension movements. We used data from 30 subjects for training/evaluation (80/20 intra-subject split).
This dataset served as the primary benchmark for fine-tuning and pre-training evaluation.

\subsubsection{Extended flexible electrode dataset (Flex-Extended dataset)}
This dataset is larger dataset with a size of approximately $4\times$ Basic Flexible Electrode Dataset, collected using the same flexible electrode setup but encompassing a wider variety of tasks including isometric contractions, dynamic movements, exoskeleton use and subjects including individuals post-stroke. Primarily used as a larger unlabeled corpus for pre-training to test scaling effects.

\subsubsection{Traditional rigid electrode dataset (Rigid-AgCl dataset)}
This dataset is collected using standard Ag/AgCl electrodes with 12 channels EMG recordings, sampled at 2000 Hz.
The amplifier and software used for acquisition were the same as both flexible electrode datasets, only the electrode type differed.
The experimental paradigms for this data also included a wide range of experiments such as grip force, button pressing, and hand motion recognition.
Its size is approximately  twice as large as that of the Basic Flexible Electrode Dataset.
We used it for cross-domain pre-training evaluation.

\subsubsection{Amputation Flexible Electrode Dataset (Flex-Amp dataset)}
This dataset contains 12-channel sEMG collected using the flexible electrode sleeves from 4 individuals with transradial amputation attempting to perform prompted finger movements, mirroring with intact limb. Ground truth labels are derived from the prompted movements. Used to evaluate model adaptability to challenging clinical populations. The details about this dataset was given in \citep{Weng2025Realtime}.

\subsection{Data Preprocessing}

Raw sEMG signals were preprocessed to improve signal quality. Bandpass filtering between 8-500 Hz was applied to remove high-frequency noise and motion artifacts while preserving the dominant EMG spectrum. Notch filters at 50 Hz and its harmonics were used to suppress powerline interference. Following the methods proposed in \citep{defossezDecodingSpeechPerception2023}, a robust scaling technique was applied, followed by clipping to minimize the impact of outliers and artifacts such as electrode movement and wire displacement.
\subsection{Baselines}

To rigorously evaluate the benefits of our proposed SSL framework, we conducted comparisons against two main categories of models:
\begin{enumerate*}
    \item state-of-the-art generic SSL frameworks adapted for sEMG, and
    \item an ablative version of our own architecture trained purely with supervision.
\end{enumerate*}
This strategy allows us to disentangle the gains from our architectural choices versus the pre-training task itself.

\textbf{Supervised-only Baseline:} This is our full SPECTRE architecture (including the CNN embedder and CyRoPE) trained from scratch on the labeled downstream task. This crucial baseline establishes the performance of our powerful, domain-adapted architecture without any self-supervised pre-training. It serves as a strong, state-of-the-art candidate model for sEMG decoding in its own right, and allows us to precisely quantify the added value of SSL.

\textbf{Generic SSL Frameworks:} We adapted several prominent SSL frameworks from other domains to serve as benchmarks for the pre-training strategy. This allows us to assess whether SPECTRE's domain-specific design choices offer tangible benefits over generic approaches.
\begin{enumerate*}
    \item \textbf{MAE \citep{heMaskedAutoencodersAre2021}:} A canonical masked signal modeling baseline. We evaluate its original form and a version enhanced with our architectural improvements (\textit{CNN MAE++}) to isolate the effect of the pre-training objective. It aims to reconstruct raw sEMG signal values in masked patches.
    \item \textbf{BIOT \citep{yang2023biot}:} A general-purpose Transformer for biosignals, which also uses a masked prediction objective.
    \item \textbf{VQ-MTM \citep{gui2024vector}:} An SSL method from the EEG domain using vector quantization on time-domain features, providing a contrast to our spectral-domain quantization.
\end{enumerate*}

\subsection{Implementation Details}

\subsubsection{Architecture settings}

For comparison, all Transformer-based models used an 18-layer encoder with a hidden dimension $d=256$ and 4 attention heads per layer. Models incorporating the CNN embedder used a 3-layer 1D CNN: Layer 1 (kernel 7, stride 2, 32 channels), GELU, MaxPool (kernel 3); Layer 2 (kernel 5, stride 1, 96 channels), GELU, MaxPool (kernel 3); Layer 3 (kernel 3, stride 1, 256 channels). This processed input sEMG segments of 100 samples (50 ms) into 256-dimensional embeddings. SPECTRE and "CNN MAE++" used RMSNorm normalization and SwiGLU activation function; other baselines used LayerNorm normalization and standard activation functions unless specified otherwise in their original papers. CyRoPE was used in SPECTRE and "CNN MAE++"; other baselines used standard learned absolute positional embeddings.

\subsubsection{Pre-training strategy}
Models employing masking used a uniform masking ratio of 30\%. The AdamW optimizer was used with a cosine learning rate schedule and a linear warm-up. The peak learning rate was $2 \times 10^{-4}$, batch size 128. Pre-training duration: 100 epochs (20 warm-up) on Flex-Basic; 20 epochs (4 warm-up) on the larger Flex-Extended and Rigid-AgCl datasets due to computational constraints.
For SPECTRE's spectral pseudo-labels, STFT used a Hann window of 64 samples (32 ms) and hop size 32 samples (16 ms). K-means clustering was performed offline on STFT vectors from the respective pre-training dataset to generate $K=500$ cluster centroids. The pre-training objective was Cross-Entropy loss.
For MAE-based models, the objective was MSE loss on the raw masked signal reconstruction.

\subsubsection{Fine-tuning strategy}
Models were fine-tuned on the training split of Flex-Basic dataset with AdamW optimizer, cosine schedule with 20 epochs warm-up, peak learning rate $1 \times 10^{-3}$, and batch size being 128. Total fine-tuning epochs was set to 60. The downstream task was predicting 5-DoF finger flexion values. The objective was MSE loss. For evaluation, we used the model checkpoint after 60 epochs.

\subsubsection{Evaluation Metrics}
Performance was primarily evaluated using Mean Absolute Error (MAE), Mean Squared Error (MSE), and Coefficient of Determination ($R^2$) between predicted and ground truth finger flexion values. There metrics were averaged across the 5 fingers and then across test samples or subjects. Lower MAE/MSE and higher $R^2$ indicate better performance. Results are reported as mean $\pm$ standard deviation across subjects or sessions where appropriate.

\subsection{Results}

\subsubsection{Effectiveness of Self-Supervised Pre-training}

\begin{table*}[!htbp]
    \centering
    \caption{Effect of self-supervised pre-training on fine finger decoding, pre-train and fine-tune on Flex-Basic dataset. Best results in bold.}
    \label{tab:performance_pretrain_intra}
    \begin{threeparttable}
        \begin{tabular}{lccccc}
            \toprule
            \textbf{Model}                                          & \textbf{Pre-trained?} & \textbf{MSE} ($\downarrow$)  & \textbf{MAE} ($\downarrow$)  & \boldmath{$R^2$} ($\uparrow$) \\
            \midrule
            \multirow{2}{*}{BIOT \citep{yang2023biot}}               & No                    & 0.0238 $\pm$ 0.0108          & 0.0896 $\pm$ 0.0259          & 0.6824 $\pm$ 0.1751           \\
                                                                    & Yes                   & 0.0227 $\pm$ 0.0130          & 0.0937 $\pm$ 0.0315          & 0.6978 $\pm$ 0.1886           \\
            \midrule
            \multirow{2}{*}{VQ-MTM \citep{gui2024vector}}            & No                    & 0.0611 $\pm$ 0.0185          & 0.1566 $\pm$ 0.0322          & 0.2249 $\pm$ 0.2287           \\
                                                                    & Yes                   & 0.0572 $\pm$ 0.0183          & 0.1486 $\pm$ 0.0299          & 0.2753 $\pm$ 0.2173           \\
            \midrule
            \multirow{2}{*}{MAE \citep{heMaskedAutoencodersAre2021}} & No                    & 0.0264 $\pm$ 0.0159          & 0.0937 $\pm$ 0.0351          & 0.6522 $\pm$ 0.2256           \\
                                                                    & Yes                   & 0.0239 $\pm$ 0.0155          & 0.0872 $\pm$ 0.0345          & 0.6807 $\pm$ 0.2274           \\
            \midrule
            + RMSNorm,                                              & No                    & 0.0214 $\pm$ 0.0122          & 0.0835 $\pm$ 0.0297          & 0.7161 $\pm$ 0.1785           \\
            SwiGLU, CyRoPE                                          & Yes                   & 0.0202 $\pm$ 0.0109          & 0.0801 $\pm$ 0.0272          & 0.7314 $\pm$ 0.1642           \\
            \midrule
            + CNN                                                   & No\textsuperscript{*} & \textbf{0.0196 $\pm$ 0.0110} & \textbf{0.0794 $\pm$ 0.0276} & \textbf{0.7380 $\pm$ 0.1636}  \\
            (CNN MAE++)                                             & Yes                   & 0.0194 $\pm$ 0.0119          & 0.0788 $\pm$ 0.0300          & 0.7425 $\pm$ 0.1709           \\
            \midrule + STFT Cluster                                 & No\textsuperscript{*} & \textbf{0.0196 $\pm$ 0.0110} & \textbf{0.0794 $\pm$ 0.0276} & \textbf{0.7380 $\pm$ 0.1636}  \\
            (SPECTRE)                                               & Yes                   & \textbf{0.0184 $\pm$ 0.0114} & \textbf{0.0770 $\pm$ 0.0287} & \textbf{0.7547 $\pm$ 0.1657}  \\
            \bottomrule
        \end{tabular}
        \begin{tablenotes}
            \footnotesize
            \item[*] SPECTRE and CNN MAE++ only differ in their pre-training strategy, their architectures are identical and their performance without pre-training is identical.
        \end{tablenotes}
    \end{threeparttable}
\end{table*}

Table \ref{tab:performance_pretrain_intra} shows the performance of comparing models trained supervisedly from scratch versus those pre-trained on Flex-Basic dataset on the test set of Flex-Basic dataset.

\begin{figure}[!htbp]
    \centering
    \includegraphics[width=\linewidth]{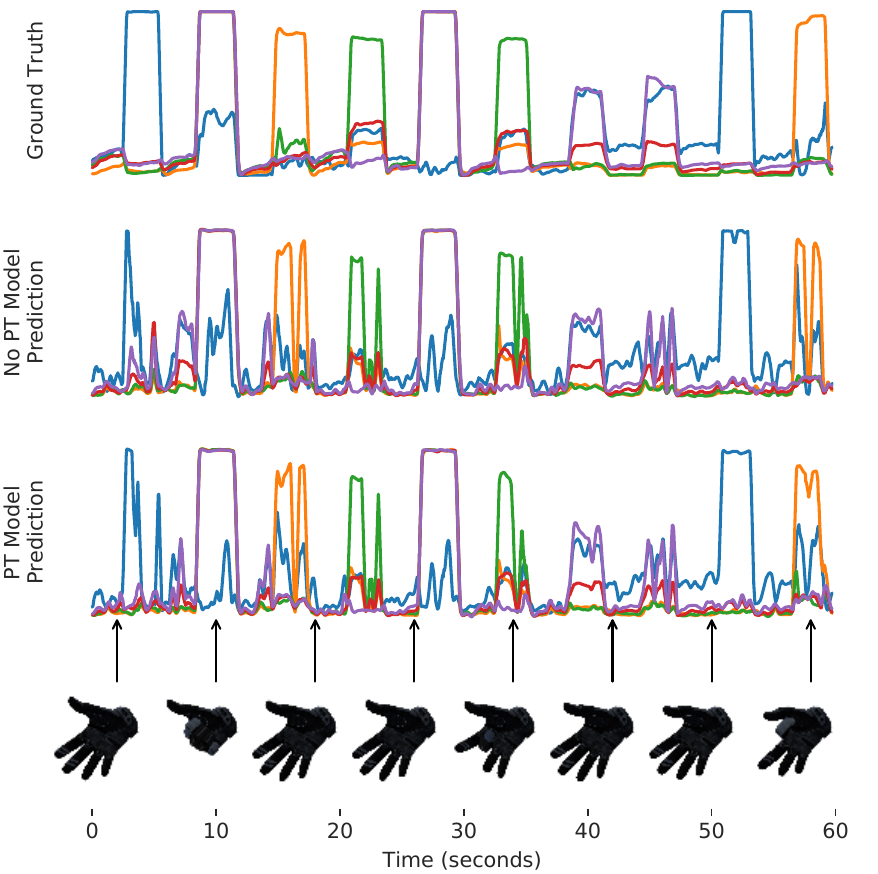}
    \caption{Decoding results for a representative non-amputated subject. Top to bottom: Ground truth finger kinematics; decoded kinematics using SPECTRE model without self-supervised pre-training; decoded kinematics using SPECTRE with self-supervised pre-training.}
    \label{fig:able-result}
\end{figure}

It can be observed from Table \ref{tab:performance_pretrain_intra} that SSL pre-training generally improves performance over supervised-only training for most architectures, e.g., MAE Base $R^2$ improves from 0.6522 to 0.6807). VQ-MTM showed poor absolute performance, potentially due to difficulties in adapting its EEG-specific design or quantization strategy effectively to sEMG. BIOT showed modest gains.
Architectural improvements (RMSNorm, SwiGLU, CyRoPE, CNN embedder) significantly boost performance even without pre-training. The SPECTRE architecture trained supervisedly ($R^2=0.7380$) substantially outperforms the basic MAE architecture trained supervisedly ($R^2=0.6522$).
Most importantly, SPECTRE's pre-training task (STFT Cluster prediction) yields the largest performance gain from pre-training compared to the MAE objective on the same architecture (CNN MAE++). SPECTRE achieves the overall best performance ($R^2=0.7547$), significantly outperforming all baselines. This highlights the benefit of the spectral pseudo-label approach combined with the optimized architecture.
To provide a qualitative illustration of this performance improvement, Figure \ref{fig:able-result} visualizes the decoding results for a representative non-amputated subject.

\begin{table*}[!htbp]
    \centering
    \caption{Effect of cross-domain pre-training. Models pre-trained on Rigid-AgCl dataset and fine-tuned on Flex-Basic. Best result in bold.}
    \label{tab:performance_pretrain_intra_trad}
    \begin{tabular}{lccc}
        \toprule
        \textbf{Model} & \textbf{MSE} ($\downarrow$)  & \textbf{MAE}  ($\downarrow$) & \boldmath{$R^2$} ($\uparrow$) \\
        \midrule
        BIOT           & 0.0231 $\pm$ 0.0117          & 0.0957 $\pm$ 0.0277          & 0.6946 $\pm$ 0.1710           \\
        MAE            & 0.0221 $\pm$ 0.0132          & 0.0847 $\pm$ 0.0315          & 0.7037 $\pm$ 0.1974           \\
        CNN MAE++      & 0.0198 $\pm$ 0.0111          & \textbf{0.0791 $\pm$ 0.0274} & 0.7374 $\pm$ 0.1607           \\
        SPECTRE        & \textbf{0.0196 $\pm$ 0.0119} & 0.0798 $\pm$ 0.0287          & \textbf{0.7389 $\pm$ 0.1744}  \\
        \bottomrule
    \end{tabular}
\end{table*}

We also investigated pre-training on the Rigid-AgCl dataset (different electrode type) and fine-tuning on Flex-Basic (Table \ref{tab:performance_pretrain_intra_trad}). Pre-training on the Rigid-AgCl dataset provided marginal benefits for the advanced architectures (CNN MAE++, SPECTRE), with performance very close to supervised-only training. MAE based baseline showed a clearer benefit compared to its supervised-only counterpart (Table \ref{tab:performance_pretrain_intra}), suggesting some knowledge transfer is possible, but the domain gap limits the gains, especially for the more optimized architectures which already perform well without pre-training. This emphasizes the importance of domain similarity for effective pre-training.

\subsubsection{Impact of Pre-training Data Scale and Domain}
We investigated how the scale and domain of the pre-training data affect SPECTRE's performance after fine-tuning on Flex-Basic dataset. Results are listed in Table \ref{tab:res_data_scale_domain}.

\begin{table*}[!htbp]
    \centering
    \caption{Effect of pre-training data source on SPECTRE Performance. Best result per metric in bold.}
    \label{tab:res_data_scale_domain}
    \begin{tabular}{lccc}
        \toprule
        \textbf{Pre-train Dataset} & \textbf{MSE} ($\downarrow$)  & \textbf{MAE} ($\downarrow$)  & \boldmath{$R^2$} ($\uparrow$) \\
        \midrule
        None                      & 0.0196 $\pm$ 0.0110          & 0.0794 $\pm$ 0.0276          & 0.7380 $\pm$ 0.1636           \\
        Flex-Basic                & 0.0184 $\pm$ 0.0114          & 0.0770 $\pm$ 0.0287          & 0.7547 $\pm$ 0.1657           \\
        Rigid-AgCl                & 0.0196 $\pm$ 0.0119          & 0.0798 $\pm$ 0.0287          & 0.7389 $\pm$ 0.1744           \\ 
        Flex-Ext                  & \textbf{0.0176 $\pm$ 0.0100} & \textbf{0.0763 $\pm$ 0.0255} & \textbf{0.7651 $\pm$ 0.1480}  \\
        \bottomrule
    \end{tabular}
\end{table*}

The results confirm the benefits of scaling: pre-training on the larger, in-domain Flex-Extended dataset yields the best performance ($R^2=0.7651$), surpassing pre-training on the smaller Flex-Basic dataset ($R^2=0.7547$). This demonstrates that SPECTRE leverages larger amounts of unlabeled data. As observed before, pre-training on the cross-domain Rigid-AgCl dataset offers negligible improvement over supervised-only training, highlighting the importance of matching pre-training and fine-tuning domains, particularly regarding sensor hardware.

\subsubsection{Performance on Amputation Data}

We evaluated the models on the challenging Amp-Flex dataset, using a self-supervised pre-train on Flex-Basic dataset and/or supervised fine-tuned on Flex-Basic dataset. Table \ref{tab:model_performance_comparison_ampute} summarizes the results, all models was evaluated after employed the few-shot calibration strategy on 1 minute data from the same individual with amputation, same as described in \cite{Weng2025Realtime}.

\begin{table*}[!htbp]
    \centering
    \caption{Performance Comparison on Individual with amputation Dataset using Pre-training and/or Fine-tuning. Best result per metric in bold.}
    \label{tab:model_performance_comparison_ampute}
    \begin{tabular}{lcccc}
        \toprule
        \textbf{Model}             & \textbf{Pre-trained?} & \textbf{Fine-tuned?} & \textbf{MSE} ($\downarrow$)  & \textbf{MAE} ($\downarrow$)  \\
        \midrule
        \multirow{3}{*}{MAE}       & No                   & Yes                 & 0.1602 $\pm$ 0.0306          & 0.2853 $\pm$ 0.0409          \\
                                   & Yes                  & No                  & 0.1617 $\pm$ 0.0426          & 0.2877 $\pm$ 0.0479          \\
                                   & Yes                  & Yes                 & 0.1530 $\pm$ 0.0312          & 0.2762 $\pm$ 0.0368          \\
        \midrule
        \multirow{3}{*}{CNN MAE++} & No                   & Yes                 & 0.1589 $\pm$ 0.0324          & 0.2926 $\pm$ 0.0378          \\ 
                                   & Yes                  & No                  & 0.1591 $\pm$ 0.0595          & 0.2931 $\pm$ 0.0532          \\
                                   & Yes                  & Yes                 & 0.1507 $\pm$ 0.0406          & 0.2751 $\pm$ 0.0452          \\
        \midrule
        \multirow{3}{*}{SPECTRE}   & No                   & Yes                 & 0.1589 $\pm$ 0.0324          & 0.2926 $\pm$ 0.0378          \\
                                   & Yes                  & No                  & 0.1574 $\pm$ 0.0418          & 0.2912 $\pm$ 0.0407          \\
                                   & Yes                  & Yes                 & \textbf{0.1349 $\pm$ 0.0336} & \textbf{0.2650 $\pm$ 0.0391} \\
        \bottomrule
    \end{tabular}
\end{table*}

\begin{figure}[!htbp]
    \centering
    \includegraphics[width=\linewidth]{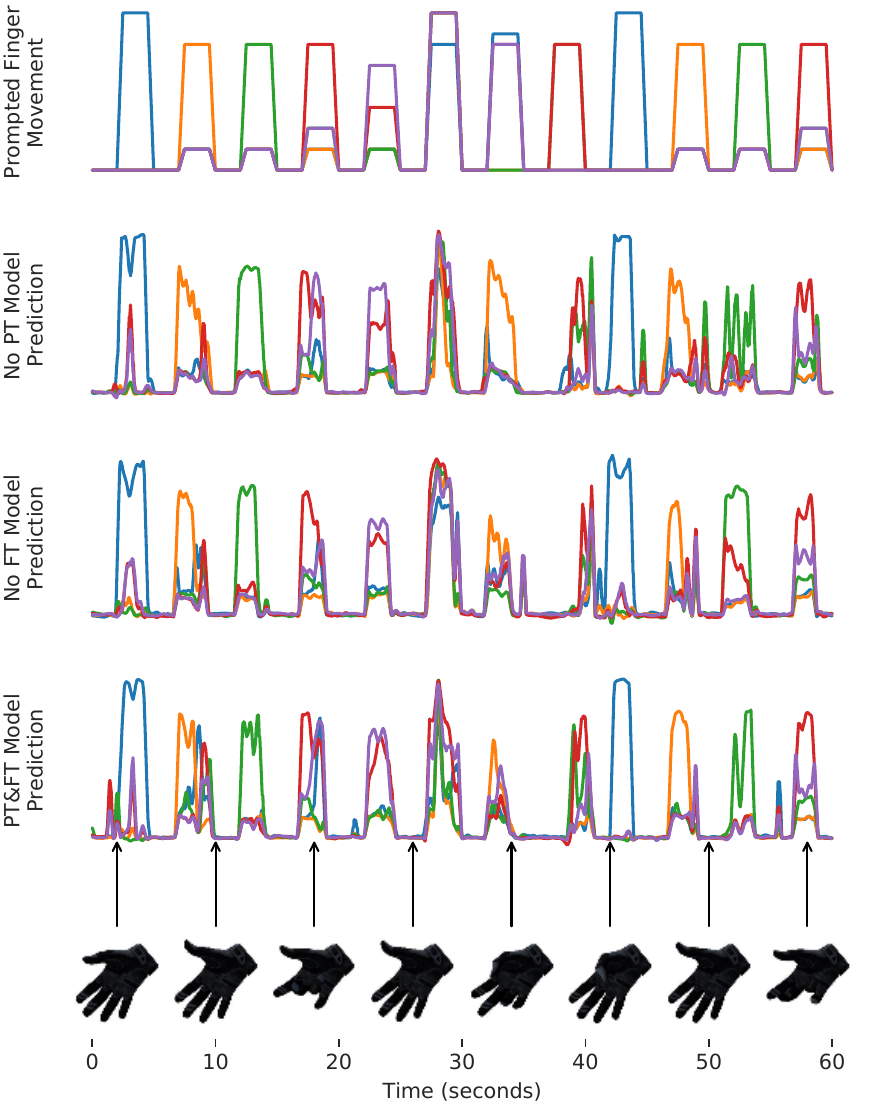}
    \caption{Decoding results for a representative individual with amputation (subject 4). Top to bottom: prompted/target kinematics (expected hand movements shown to the subject), against which the decoded movements are compared; decoded kinematics using SPECTRE model with finetuning on labeled dataset; decoded kinematics using SPECTRE model with self-supervised pre-training; decoded kinematics using SPECTRE model with self-supervised pre-training and fine-tuning.}
    \label{fig:ampute-result}
\end{figure}

For all models, the combination of self-supervised pre-training and supervised fine-tuned on data from individuals without amputation and few-shot calibration on amputation-specific data yields the best performance.
SPECTRE, when pre-trained, fine-tuned and calibrated, achieves significantly lower MAE and MSE compared to all other configurations and baseline models. This demonstrates its ability to learn generalizable sEMG features during pre-training that can be effectively adapted to the unique and often weaker or altered muscle activation patterns present in individuals with amputation.
Relying solely on self-supervised pre-training without supervised fine-tuning, or only on supervised fine-tuning without leveraging self-supervised knowledge, results in substantially worse performance, highlighting the synergistic effect of the two stages for adapting to challenging target populations.
Figure \ref{fig:ampute-result} provides a compelling qualitative view of SPECTRE's effectiveness on the challenging individual with amputation dataset. The plots compare the target kinematics against the decoded outputs from different training configurations.

\subsubsection{Ablation Studies}
To validate the contributions of SPECTRE's key components, we performed ablation studies, focusing on CyRoPE and the STFT pseudo-label target. Results are shown in Table \ref{tab:ablation}.

\begin{table*}[htbp]
    \centering
    \begin{threeparttable}
        \caption{Ablation study for SPECTRE. Best result per metric in bold.}
        \label{tab:ablation}
        \begin{tabular}{lcccccccc}
            \toprule
            \textbf{Pre-train Target} & \textbf{PE Type} & \textbf{Pre-trained?} & \textbf{MSE}   ($\downarrow$) & \textbf{MAE} ($\downarrow$)  & \boldmath{$R^2$} ($\uparrow$) \\
            \midrule
            STFT Clusters             & Absolute PE      & No                   & 0.0219 $\pm$ 0.0140           & 0.0835 $\pm$ 0.0314          & 0.7091 $\pm$ 0.1985           \\
            STFT Clusters             & Absolute PE      & Yes                  & 0.0206 $\pm$ 0.0123           & 0.0812 $\pm$ 0.0293          & 0.7252 $\pm$ 0.1804           \\
            \midrule
            Raw Signal Clusters       & CyRoPE           & Yes                  & 0.0189 $\pm$ 0.0117           & 0.0777 $\pm$ 0.0297          & 0.7469 $\pm$ 0.1746           \\
            \midrule
            STFT Clusters             & CyRoPE           & No                   & 0.0196 $\pm$ 0.0110           & 0.0794 $\pm$ 0.0276          & 0.7380 $\pm$ 0.1636           \\
            STFT Clusters             & CyRoPE           & Yes                  & \textbf{0.0184 $\pm$ 0.0114}  & \textbf{0.0770 $\pm$ 0.0287} & \textbf{0.7547 $\pm$ 0.1657}  \\
            \bottomrule
        \end{tabular}
    \end{threeparttable}
\end{table*}

To evaluate the effectiveness of CyRoPE, this study first compared the performance of models using traditional absolute PE versus CyRoPE. Keeping other components unchanged, only the position embedding method of the SPECTRE model was replaced. Results showed models with absolute PE performed worse on all metrics after pre-training compared to the model using CyRoPE. Specifically, the $R^2$ value dropped from 0.7547 to 0.7252. Furthermore, even without pre-training, it still lagged behind the model using CyRoPE by 3.92\% in the $R^2$ metric. This indicates that CyRoPE can more effectively capture the spatiotemporal features of sEMG signals, thereby enhancing the model's decoding performance.

To validate the effectiveness of spectral embedding pseudo-labels, this study compared the performance of models using STFT clustering pseudo-labels versus simply clustering the flattented raw signals. Experimental results showed that the model using flattened EMG data for pseudo-label generation had slightly lower performance metrics ($R^2= 0.7469$) compared to the model using STFT clustering ($R^2=0.7547$). Although the simple flatten strategy achieved some performance improvement compared to directly reconstructing the raw signal ($R^2= 0.7429$), STFT clustering pseudo-labels enabled the model to learn superior feature representations. This is likely because its ability to guide the model to focus on more physiologically meaningful frequency components within the sEMG signal.

The ablation study results clearly demonstrate that both CyRoPE and spectral pre-training are key components contributing to the performance improvement of the SPECTRE model, enabling more effective utilization of the information gain from self-supervised pre-training. This also suggests that STFT clustering pseudo-labels represent a highly efficient characterization of myoelectric features, and their combination with pre-training significantly enhances the model's decoding capabilities.

\section{Discussion} \label{sec:discussion}

The results support the efficacy of the SPECTRE framework. Our central hypothesis—that a domain-specific SSL approach, which leverages spectral features and sensor topology, would outperform generic methods—is substantiated. The superior performance stems from the synergy between our physiologically-grounded pre-training strategy and the topology-aware CyRoPE, built on a robust CNN-Transformer architecture.

\subsection{The Advantage of Spectral Pre-training}
The ablation study (Table \ref{tab:ablation}) clearly indicates that predicting clustered STFT pseudo-labels is more effective than reconstructing raw signals (MAE objective) or clustered raw signals. sEMG signals are inherently noisy; reconstructing the raw signal forces the model to potentially overfit to noise patterns. In contrast, the STFT captures the power distribution across frequencies, which is physiologically more stable and informative regarding muscle activation levels and fiber recruitment \citep{4121265}. Clustering these spectral representations further provides a canonical, denoised target, guiding the model to learn robust features invariant to certain types of noise and minor signal variations. This aligns with successes in speech recognition where models learn from quantized acoustic features (like MFCCs or learned units) rather than raw audio \citep{hsuHuBERTSelfSupervisedSpeech2021}. SPECTRE adapts this principle to the specific characteristics of sEMG.

\subsection{Explicit Spatio-Temporal Modeling via CyRoPE}
The significant performance drop when replacing CyRoPE with standard absolute PE underscores the importance of appropriately modeling the structure of multi-channel sEMG. Finger movements result from the coordinated activity of multiple muscles, whose signals are captured across spatially distributed electrodes, often in an annular configuration around the forearm. CyRoPE explicitly encodes this topology by factorizing temporal (linear) and spatial (annular) position information. The annular encoding allows the self-attention mechanism to readily learn relationships between adjacent and opposing muscle groups based on their relative angular position, complementing the learning of temporal dependencies. This geometric inductive bias provided by CyRoPE appears highly beneficial for decoding complex muscle synergies underlying fine motor control.

\subsection{Architecture Synergies}
The combination of the channel-independent CNN embedder, the advanced Transformer backbone (with RMSNorm, SwiGLU), and CyRoPE creates a powerful architecture even without pre-training (Table \ref{tab:performance_pretrain_intra}). The CNN effectively handles local feature extraction and patching, while the Transformer, enhanced by CyRoPE, models global spatio-temporal interactions. The improvements from RMSNorm and SwiGLU, while perhaps incremental individually, contribute to overall stability and performance on the challenging sEMG data.

\subsection{Data Scale and Domain Specificity}
Our results confirm that, like other deep learning domains, SSL for sEMG benefits from larger datasets (Table \ref{tab:res_data_scale_domain}). However, the strong dependence on domain similarity (flexible vs. rigid electrodes) highlights a key challenge. Differences in electrode hardware significantly alter signal characteristics (e.g., impedance, spatial filtering). While SPECTRE's spectral pre-training shows some robustness, achieving optimal performance likely requires either large-scale pre-training on highly diverse data encompassing various hardware types or the development of more sophisticated domain adaptation/generalization techniques tailored for sEMG \citep{zhang2022domain}.

\subsection{Clinical Relevance for Individuals with Amputations}
The strong performance of pre-trained and fine-tuned SPECTRE on the amputation dataset (Table \ref{tab:model_performance_comparison_ampute}) is particularly encouraging. The sEMG signals of individuals with amputation often differ significantly from those of individuals without amputation due to residual limb physiology, nerve regeneration, and altered motor control strategies \citep{sarrocaMuscleActivationGait2021}. SPECTRE's ability to leverage knowledge learned from individuals without amputation data and adapt it effectively with minimal calibration data demonstrates its potential for practical prosthetic control systems, where acquiring extensive labeled data from each user is often infeasible. The spectral pre-training likely helps capture fundamental muscle activation patterns that persist even after amputation.

\subsection{Limitations and Future Work}
Although the proposed SPECTRE achieves promising results, some limitations may exist.
\begin{enumerate*}
    \item \textit{Data Diversity:} While using multiple datasets, the pre-training corpora could be further expanded to include more subjects, tasks, and recording conditions to enhance generalization. Integrating publicly available large-scale sEMG datasets \citep{salterEmg2poseLargeDiverse2024, sivakumarEmg2qwertyLargeDataset2024} is a key next step.

    \item \textit{Real-world Robustness:} In current study, the experiments were conducted on data collected in relatively controlled settings. Performance in real-world scenarios with more pronounced artifacts, electrode shifts, and varying fatigue levels needs further evaluation. Online adaptation mechanisms might be necessary.

    \item \textit{CyRoPE Assumptions:} CyRoPE assumes a roughly uniform annular electrode arrangement. Its performance with highly irregular or non-annular placements needs investigation. More flexible spatial encoding methods could be explored.

    \item  \textit{Computational Cost:} While the SPECTRE is effective to deocode fine-grained movement, the Transformer architecture can be computationally demanding. Exploring model compression techniques (pruning, quantization) or more efficient Transformer variants \citep{kitaev2020reformer} is important for deployment on wearable devices.
    \item  \textit{Beyond Kinematics:} This work focused on kinematic decoding. Extending SPECTRE to decode kinetic information (e.g., force levels \citep{fangSimultaneousSEMGRecognition2022}) simultaneously is a valuable future direction.
    \item  \textit{Interpretability:} Understanding the specific spectral features and spatio-temporal patterns learned by SPECTRE could provide neurophysiological insights and guide further model improvements.
\end{enumerate*}

\section{Conclusion} \label{sec:conclusion}
In this work, we introduced SPECTRE, a novel self-supervised learning framework specifically engineered to address the core challenges of fine-grained sEMG decoding. We demonstrated that generic SSL methods fall short due to their inability to handle the high noise levels and ignore the critical sensor topology of sEMG. SPECTRE overcomes these issues through two targeted innovations: a pre-training task on clustered spectral embeddings that focuses on physiologically relevant, noise-robust features, and Cylindrical Rotary Position Embedding (CyRoPE), which injects essential geometric inductive bias about the sensor layout into the model.

Our comprehensive experiments show that SPECTRE significantly outperforms both strong supervised baselines and generic SSL approaches in the demanding task of continuous fine finger movement decoding. The findings underscore a critical insight for the field: the future of representation learning for biosignals lies not in the direct application of generic frameworks, but in the thoughtful design of domain-specific solutions that respect the underlying physiology and physics of the signal acquisition process. While we have outlined avenues for future work to enhance robustness and generality, SPECTRE provides a powerful new baseline and a promising foundation for the development of the next generation of practical and effective myoelectric interfaces.

\section*{Acknowledgements}

This work was supported by the Brain Science and Brain-like Intelligence Technology-National Science and Technology Major Project [grant numbers 2022ZD0208901]; the National Natural Science Foundation of China [grant numbers W2411084, 82372084]; the Sichuan Science and Technology Program [grant number 2025YFHZ0261]; the Key R\&D projects of the Science\&Technology Department of Chengdu [grant number 2024-YF08-00072-GX]; the China Postdoctoral Science Foundation [grant number 2024M760359]; and the Postdoctoral Fellowship Program of CPSF [grant number GZC20240211].


\bibliographystyle{abbrv}
\bibliography{reference} 

\end{document}